\documentstyle[12pt]{article}
\global\arraycolsep=2pt 
\begin{document}
\def\HQQET{HQ$\overline{\rm Q}$ET}
\makeatletter
\def\fmslash{\@ifnextchar[{\fmsl@sh}{\fmsl@sh[0mu]}}
\def\fmsl@sh[#1]#2{%
  \mathchoice
    {\@fmsl@sh\displaystyle{#1}{#2}}%
    {\@fmsl@sh\textstyle{#1}{#2}}%
    {\@fmsl@sh\scriptstyle{#1}{#2}}%
    {\@fmsl@sh\scriptscriptstyle{#1}{#2}}}
\def\@fmsl@sh#1#2#3{\m@th\ooalign{$\hfil#1\mkern#2/\hfil$\crcr$#1#3$}}
\makeatother
%
\thispagestyle{empty}
\begin{titlepage}

\begin{flushright}
CERN-TH.7523/94 \\
TTP 94--31 \\
\end{flushright}

\vspace{0.3cm}

\begin{center}
\Large \bf HQ$\overline{\mbox{\bf Q}}$ET: An Effective Theory Approach
to \\Heavy Quarkonia Decays
\end{center}

\vspace{0.8cm}

\begin{center}
{\large Thomas Mannel} \\
{\sl Institut f\"{u}r Theoretische Teilchenphysik,
     D -- 76128 Karlsruhe, Germany.} \vspace*{5mm} \\
{\large Gerhard A.\ Schuler}\footnote{Heisenberg Fellow.} \\
{\sl Theory Division, CERN, CH-1211 Geneva 23, Switzerland}
\end{center}

\vspace{\fill}

\begin{abstract}
\noindent
We discuss systems containing a heavy quark and a heavy
antiquark in the infinite mass limit of QCD. Studying the
limit of equal velocities for both heavy particles, we
formulate an effective theory approach to heavy quarkonia-like
systems. The method is well suited to processes in which
the two heavy quarks annihilate, such as electromagnetic
and strong decays of charmonium and bottomonium and weak
decays of $B_c$.
\end{abstract}
\vfill
\centerline{(Submitted to Phys.\ Lett.\ B)}
\bigskip\bigskip
\vspace{0.8cm}
\noindent
CERN-TH.7523/94 \\
TTP 94--31 \\
December  1994
\end{titlepage}

\section{Introduction}
In the past five years considerable progress has been made in heavy
quark physics by studying systems with a single heavy quark in the
infinite mass limit of QCD \cite{SV87}. The  mass $m$
of the heavy quark sets a
scale large compared to the intrinsic scale $\Lambda_{QCD}$ of the
light QCD-degrees of freedom, and the appearance of such disparate mass
scales allows us to use an effective theory treatment of systems with
a single heavy quark. This effective theory, the so-called heavy quark
effective theory (HQET) has the interesting property of additional
symmetries which are not present in full QCD. These
symmetries, in addition to the usual machinery of effective theory, are a
powerful tool in heavy hadron physics, which allows us to make QCD
based and in many cases even model independent statements. The progress
in this field is well documented in more or less extensive review
articles \cite{reviews}.

Almost all the applications considered so far deal with the one
heavy particle sector of QCD. In HQET particle and antiparticle number
are separately conserved and all applications deal with either a single
heavy quark or a single heavy antiquark. First attempts to consider
states containing two heavy (anti)quarks or a quark and an antiquark
revealed some difficulties \cite{GK91} when one calculates
QCD radiative corrections. The anomalous dimensions of operators having
matrix elements with states containing two heavy quarks turn out to be
complex, at least if interpreted in the na\"\i ve way. In addition,
the imaginary parts behave as $1/v$, where $v$ is the relative
three velocity of the heavy quarks.
Subsequent
investigations \cite{KM93} showed that the
imaginary parts of the anomalous
dimensions yield phase factors which have to be interpreted as
the non-abelian analogue of the Coulomb phase well known from
electrodynamics. These phases are an infrared contribution, which has to
be absorbed into the states. After redefinition
of the states the anomalous
dimensions are real, well behaved as the relative velocity becomes small,
and hence are the true short distance contribution.

Heavy quarkonia have to show up in the sector of HQET containing
a heavy quark and a heavy antiquark. The velocities of the two heavy
particles in such a state differ only by a small amount of the order
$\Lambda_{QCD} / m$, and hence one wants to describe heavy quarkonia in
the limit, where the two heavy quarks move with the same velocity, which
is then identified with the velocity of the heavy quarkonium. However,
such a limit is ill defined for static quarks due to the divergent
phases mentioned above. A proper treatment of these phases lies at the
heart of the formulation  of a Heavy Quarkonium Effective
Theory (\HQQET), since these phases are related to the binding mechanism
of the heavy quarkonium states. We shall see in what follows that
\HQQET{} may not simply be related to the two particle sector
of HQET, because the divergent phases
prevent us from taking the strict infinite mass limit.

In this paper we shall clarify the underlying assumptions necessary
to formulate \HQQET{} based on the static limit. In the next section
we shall give a  theoretical
description of the method, which has been
applied to inclusive heavy quarkonia decays already in
\cite{MS94} and compare to related ideas of Bodwin, Braaten
and Lepage (BBL) \cite{BB94}. In section 3 we discuss how
to calculate QCD and higher order non-perturbative corrections in
\HQQET .

\section{Formulation of HQ$\overline{\mbox{Q}}$ET}
Our starting point is the Lagrangian of QCD, which we expand in inverse
powers of $1/m$ \cite{MRRderivation,KT91}. The
part of the Lagrangian involving the heavy quark is unique up to terms of
order $1/m$ and is given by
\begin{equation}
{\cal L} = \bar{h}^{(+)}_v (iv D) h^{(+)}_v +
\left( \frac{1}{2m} \right) \bar{h}^{(+)}_v  i \fmslash{D}
P_- i \fmslash{D}  h^{(+)}_v  + {\cal O} (1/m^2) ,
\label{lexp}
\end{equation}
while the corresponding expansion for the field $Q$ of full QCD
is given by
\begin{equation} \label{>>>}
Q = \exp (-im\,vx)\left( 1 + \frac{1}{2m} P_- \fmslash{D}
               + {\cal O} (1/m^2) \right) h^{(+)} (x) ,
\end{equation}
where $P_\pm = (1\pm \fmslash{v})/2$.
The superscript $(+)$ indicates that the field $h_v^{(+)}$ describes
a static quark moving with the velocity $v$; correspondingly we introduce
also the field $h_w^{(-)}$, which describes an antiquark moving with the
velocity $w$. The Lagrangian and the expansion of the full QCD field for
the antiquark field is obtained from (\ref{lexp}) and (\ref{>>>}) by the
replacement $v \to - w$.

Let us first consider only the static, mass independent term
of the expansions
(\ref{lexp}) and (\ref{>>>}) and write the Lagrangian for a
two-particle system
consisting of a static quark and a static antiquark as
\begin{equation} \label{lstat}
{\cal L} = \bar{h}_v^{(+)} (ivD) h_v^{(+)} -
           \bar{h}_w^{(-)} (iwD) h_w^{(-)}
\ .
\end{equation}
Based on such a Lagrangian we may now consider the matrix
elements involving
two heavy particle states; as an example we shall study
\begin{equation} \label{fullME}
G = \langle A | \bar{Q} (x) \Gamma Q (x) | 0 \rangle ,
\end{equation}
where $A$ is a state containing a heavy quark and a
heavy antiquark moving with
velocities $v$ and $w$ respectively.
In the static limit this matrix element becomes
\begin{equation} \label{mestat}
G_{\mbox{static}} = \widetilde{\langle A |} \bar{h}_v^{(+)} (x)
\Gamma h_w^{(-)} (x) | 0 \rangle \exp [im(v+w)x] ,
\end{equation}
where the tilde denotes the static limit of the state.
Logarithmic dependences on the heavy quark mass may be calculated
using renormalization group improved perturbation theory in the
framework of HQET. The one-loop QCD radiative corrections
to a current of this type have been calculated in \cite{GK91}
and \cite{KMM93}.
It turns out that in a na\"\i ve calculation the anomalous
dimension seem to acquire an imaginary part, which for $v \to w$ develops
a divergence of the general structure
\begin{equation} \label{imag}
\mbox{Im }\gamma = f(\alpha_s)
                   \frac{1}{\sqrt{(vw)^2 - 1}}
\end{equation}
where $f(\alpha_s)$ is known up to two loops \cite{KMM93}
\begin{equation}
f(\alpha_s) = \frac{4}{3} \alpha_s \left(
1 + \frac{\alpha_s}{4 \pi} \left[ \frac{31}{3} -
\frac{10}{9} n_f \right]
+ \cdots \right)
\ .
\end{equation}
The real part of the anomalous dimension vanishes as $v \to w$ due to
current conservation, and
the solution of the renormalization group equation with a purely imaginary
anomalous dimension yields for $v \to w$ a phase factor of the form
\begin{equation} \label{phase}
\exp (i \phi (vw )) =
\exp\left\{ i \frac{1}{\sqrt{(vw)^2 - 1}}
\int\limits_{\alpha_s (m)}^{\alpha_s (\mu)} d \alpha
    \frac{f(\alpha)}{\beta(\alpha)} \right\} \quad ; \quad
\beta (\alpha (\mu )) = \mu \frac{\partial}{\partial \mu} \alpha( \mu )
\end{equation}
which is ill behaved in the limit $v \to w$.

This divergence of the imaginary part prevents us from taking the
limit $v \to w$ for two heavy static particles. On the other hand,
this is exactly the limit in which we want to describe heavy quarkonia
states. From this we conclude that the purely
static Lagrangian (\ref{lstat})
in not appropriate for the description of heavy quarkonia states.

In fact, the phases are the non-abelian counterpart of the Coulomb phases
well known in QED. They are related to the long range part of the one
gluon (one photon) exchange potential, which decreases too slowly and
thus leads to infrared problems.
Consequently, these phases are an infrared
effect and have to be absorbed  into the states \cite{KM93}.

In the channels, where this potential is attractive,
bound states may occur,
and these phases are directly related to the binding mechanism. This may
be explicitly seen for the abelian case using eikonal methods,
which correspond
to the heavy mass limit \cite{fried}. The binding of a heavy
quarkonium is
clearly an infrared effect and has to be reproduced by the dynamics of a
properly constructed effective theory for quarkonia. In such a state
the two velocities differ
only by a small amount of order $1/m$ which is a hint that we need to
go beyond the static limit to describe quarkonia states.

In order to see how higher-order terms of the Lagrangian cure the
problem, we make use of the fact that the full QCD Lagrangian is
independent of the arbitrarily chosen velocity vectors $v$ and $w$
\cite{repara}. The only kinematic quantity entering in full QCD are
the true momenta of the particle $p = m v + k$ and the antiparticle
$p' = m w -  k'$, where $k$ and $k'$ are the residual momenta of the
two heavy quarks. Thus we are led to define
\begin{equation}
{\cal V} = v + \frac{\stackrel{\longleftarrow}{iD}}{m}
\mbox{ and }  {\cal W} = w + \frac{iD}{m}
\end{equation}
corresponding to $p/m$ and $p'/m$ respectively. These combinations are
invariant under infinitesimal reparametrizations of the velocities
$v \to v + \delta v$ and $w \to w + \delta w$
\cite{repara}, since under such a reparametrization we have
$D \to D - m \delta v$ for the quark and
$D \to D - m \delta w$ for the antiquark moving with the velocity $-w$.
In a full QCD calculation the singularity corresponding
to the one appearing in (\ref{imag}) occurs in the imaginary part of
the vertex function as
\begin{equation}
\mbox{Im } \Gamma (p, p') = f(\alpha_s)
                   \frac{m^2}{\sqrt{(pp')^2 - m^4}}
                  \ln \left( \frac{m}{\lambda} \right) ,
\end{equation}
where $\lambda$ now is an infrared regulator, revealing the
infrared origin of the singularity. The vertex function of full
QCD depends on the full quark momenta $p$ and $p'$, which are
split into a
large piece $m v$ ($m w$) and a residual part $k$ ($k'$) as we switch
to the effective theory. This is, however, an artificial procedure, and
the singularity may be reproduced either in the
dependence on the velocity
or through the residual momenta.

In heavy quarkonia the velocities of its heavy constituents are
almost equal, $vw \sim 1$, and one would rather go to the limit
$v = w$ and reproduce the divergence of the imaginary part as a
singularity in the dependence on the
residual momenta. This is achieved formally by reinserting the
full momenta for the velocities in the divergent phase,
i.e.~by the replacement $v \to {\cal V}$
and $w \to {\cal W}$, and we obtain from (\ref{phase})
\begin{equation}
\exp (i \phi (vw )) \bar{h}_v^{(+)} \Gamma h_{-w}^{(-)}
\quad \longrightarrow \quad
\bar{h}_v^{(+)}
\exp (i \phi ({\cal V} {\cal W}))
\Gamma h_{-w}^{(-)}
\ .
\end{equation}
If we now consider the limit  $v \to w$ we have also
${\cal V} \to {\cal W}$, but now the phase depends on the residual
momenta, which are represented by the covariant derivatives acting
on the heavy quark fields
\begin{equation}
\exp (i \phi ({\cal V} {\cal W})) \to \exp (i \phi ({\cal V}^2 ))
= \exp\left\{ i \frac{1}{\sqrt{ {\cal V}^2 - 1}}
\int\limits_{\alpha_s (m)}^{\alpha_s (\mu)} d \alpha
    \frac{f(\alpha)}{\beta\alpha} \right\}
\ ,
\end{equation}
which means that we have rewritten the singular phases in such a
way that they now depend on the difference of the residual momenta
rather than on the difference of the velocities.

However, the difference between $v$ and ${\cal V}$ is a term of
higher order in $1/m$, which was added in such a way that
${\cal V}$ is a reparametrization invariant quantity. In order
to construct a leading order Lagrangian capable of reproducing
the (infrared) phase factors and also generate binding of the
two heavy quarks, we rewrite the static Lagrangian (\ref{lstat})
in a reparametrization-invariant form;
in this way we obtain as the leading order Lagrangian
\begin{eqnarray}
{\cal L}_0 &=& \frac{m}{2}\bar{h}_v^{(+)} ({\cal V}^2 - 1) h^{(+)}_v
+ \frac{m}{2}\bar{h}_{-w}^{(-)} ({\cal W}^2 - 1) h^{(-)}_{-w}
\nonumber\\
           &=& \bar{h}_v^{(+)} (ivD) h^{(+)}_v
             + \bar{h}_v^{(+)} \frac{(iD)^2}{2m} h^{(+)}_v
             - \bar{h}_w^{(-)} (iwD) h^{(-)}_w
             + \bar{h}_w^{(-)} \frac{(iD)^2}{2m} h^{(-)}_w
\ ,
\label{lnull}
\end{eqnarray}
where we have replaced $w \to -w$ in the last step. In the Lagrangian
(\ref{lnull}) we now may perform the limit $v \to w$ without encountering
a problem in the calculation of QCD radiative corrections;
we now have already
to leading order the scale of the heavy mass $m$, and ultraviolet
contributions
show up as $\ln (\Lambda / m)$ ($\Lambda$ being the
ultraviolet cut-off) and
infrared contributions as $\ln (m / \lambda)$
($\lambda$ being the infrared cut-off).
In this way a clean separation between (calculable)
short distance effects and
(non-perturbative) infrared contributions is achieved.
The diverging phases now
show up as a singularity in the residual relative momentum,
hence as an infrared effect.

We also expect that the Lagrangian (\ref{lnull})
ensures the existence of bound states, corresponding
to the ``unperturbed''
heavy quarkonia states. However, this also shows that -- unlike for
heavy-light systems -- there is no infinite mass limit for quarkonia; the
``unperturbed'' states described by (\ref{lnull}) will still be mass
dependent.

The Lagrangian (\ref{lnull}) is the minimum that is needed to shift the
phases from the velocity dependence into the residual momenta. The spin
dependent terms appearing as well in order $1/m$ do not contribute to the
infrared behaviour. This is well known from QED, where all
infrared contributions
are independent of the spin of the radiating particle;
only the total charge
plays a role. Spin symmetry thus remains unbroken to leading order, and
spin symmetry breaking effects may be calculated as perturbations without
encountering infrared problems.

The leading order Lagrangian (\ref{lnull}) resembles very much the one of
non-relativistic QCD (NRQCD) as formulated by
Caswell and Lepage \cite{CL86}.
They suggest an expansion in $v/c$, where $v$ is
the typical relative velocity
of the two heavy quarks bound in the quarkonium.
Although it does not seem
that the two approaches are completely equivalent \cite{MS94}, they have
many common features.

The binding of the two heavy quarks will generate
a small non-perturbative
scale $\tilde\Lambda$, which now in general depends on the heavy mass.
Although the bottomonium and the charmonium are
not Coulombic systems,
the case of a $\alpha / r$ potential is instructive.
Neglecting any running of
$\alpha$, the size of such a  Coulombic system is
$R_{Bohr}= 1/(\alpha m)$, which
is large compared to the Compton wavelength $\lambda_Q = 1/m$.
However, the
small scale $1/R_{Bohr}$ depends on the mass such that it does not
approach a finite limit as $m \to \infty$, even for running $\alpha_s$.

The Lagrangian (\ref{lnull}) is the starting point of an effective theory
treatment of heavy quarkonium decays. The processes
which may be considered
in this type of approach are decays of heavy quarkonia,
in which the two heavy
quarks annihilate.
The annihilation process is governed by a large energy  scale
set by the heavy quark mass $m$, while the binding
of the quarkonium introduces
a small scale $\tilde\Lambda$.
The appearance of these disparate mass scales
allows for an effective theory treatment, yielding
an expansion in powers of $\tilde\Lambda / m$
of the relevant amplitudes of full QCD.

Heavy Quark Spin symmetry implies that the ``unperturbed''
heavy quarkonia states
fall into degenerate spin symmetry quartets. For a given orbital angular
momentum $\ell$ and
radial excitation quantum number $n$, the four states
(in the spectroscopic notation ${}^{2S+1}\ell_J$)
\begin{equation}
[n{}^1 \ell_\ell \quad n{}^3 \ell_{\ell-1} \quad n{}^3 \ell_\ell \quad
 n{}^3 \ell_{\ell+1} ]
\label{fourstates}
\end{equation}
form such a spin symmetry quartet. An exception are the $S$ waves
($\ell = 0$), for which the three polarization directions of the
$n{}^3S_1$ and the $n{}^1S_0$ form the spin symmetry quartet.
The consequences of this symmetry for transitions
from an excited quarkonium to the ground states have been investigated
recently \cite{CW94}.

In order to exploit the consequences of the spin symmetry for
the transition matrix elements
we shall use the trace formalism. We denote with $|Y_\ell\rangle$
the spin symmetry quartet consisting of the spin singlet and the
spin triplet for a given orbital angular momentum $\ell$. The
coupling of the heavy-quark spins may be represented by the
matrices
\begin{equation}
H_Y (v)= \left\{ \begin{array}{l}
      P_+ \gamma_5 \mbox{ for the spin singlet} \\
      P_+ \fmslash{\epsilon} \mbox{ for the spin triplet}
              \end{array} \right.
\ .
\end{equation}
Using these representations one may then analyse the spin structure
of matrix elements for processes involving quarkonia. As a simple example
we study a matrix element like (\ref{fullME}) with $|A \rangle$ being
a quarkonium state $| Y (J,J_z,n,l) \rangle$. In the heavy mass limit
we have
\begin{equation} \label{Ystat}
\langle \widetilde{Y} (J,J_z,n,l,S) | \bar{h}_v^{(+)}
\Gamma h_w^{(-)} | 0 \rangle = a(n,l) \mbox{Tr } \{\bar{H}_Y \Gamma \}
\end{equation}
where $a(n,l)$ is independent of the spin coupling of the heavy quarks.
A simple consequence of spin symmetry is that the basis of the 16 Dirac
matrices may be reduced to four, which are the generalization of the
Pauli matrices \cite{MS94,Ma94}. Furthermore, from (\ref{Ystat})
we have for the ground state quarkonia
\begin{equation}
\langle \widetilde{Y} (1,J_z,0,0,1) | \bar{h}_v^{(+)}
\fmslash{\epsilon} (J_z) h_w^{(-)} | 0 \rangle
= \langle \widetilde{Y}  (0,0,0,0,0) | \bar{h}_v^{(+)}
\gamma_5  h_w^{(-)} | 0 \rangle .
\end{equation}
In a simple wave function picture this means that the wave functions
at the origin of the two ground state quarkonia are equal in the heavy
mass limit.

\boldmath
\section{Higher Order Correction in \HQQET}
\unboldmath
In a similar way as in HQET the corrections in \HQQET{}
fall into two classes.
The first type of corrections are the QCD radiative
corrections, which in general
may be calculated systematically.
The second class are the recoil corrections
appearing as a power series in $\tilde\Lambda / m$; they originate from
matrix elements of higher dimensional operators induced by the expansion
of the currents and the Lagrangian.

Let us first consider the QCD radiative corrections,
which may be calculated
systematically using the Feynman rules of the effective theory.
Based on our choice for the leading order term ${\cal L}_0$,
which is the sum of ${\cal L}_{\mbox{static}}$ and the first
order term $\bar{h} (iD)^2 h / (2m)$, one derives for
the propagator of the heavy particle with
velocity $v$ and residual momentum $k$
\begin{equation} \label{propeff}
H (k) = P_+ \frac{i}{vk + \frac{1}{2 m} k^2 + i \epsilon} ;
\end{equation}
the corresponding expression for the antiparticle is obtained by
the replacement $v \to - v$. This expression contains
all orders in $1/m$;
however, the reason why we
had to include these higher-order terms was that this removes the
divergent phase occurring in the limit  of small relative velocity.
This phase may be absorbed as a long-distance effect into the
states, which thus have to evolve according to the dynamics
dictated by ${\cal L}_0$ given in (\ref{lnull}). The ``true''
short distance contribution is well behaved as the
relative velocity of the
heavy particles vanishes. Furthermore, it may be expanded again in
powers of $(1/m)^n$ without encountering a problem. This is at least true
at the one-loop level, where (\ref{propeff}) is the propagator of NRQCD
\cite{CL86}, and it has been shown in \cite{BB94}
that the coefficients of
the ultraviolet divergences at the one-loop level may be expanded in
$1/m$.

On the other hand, if we start directly from the static limit using the
propagator of HQET
\begin{equation} \label{propstat}
H_{stat} (k) = P_+ \frac{i}{vk + i \epsilon}  ,
\end{equation}
imaginary parts will show up, which become ill-defined as velocities
of heavy quarks coincide, such as
\begin{equation}
\int \frac{d^4 k}{(2 \pi)^4} \frac{1}{k^2} \delta (vk) \delta (v'k)
\to \int \frac{d^4 k}{(2 \pi)^4} \frac{1}{k^2} (\delta (vk))^2
\mbox{ as } v \to v '
\end{equation}
at the one-loop level, which are contributions to the
divergent phase. However, these are absorbed into the states as being a
long distance contribution which is generated by the infrared dynamics
of ${\cal L}_0$ and consequently they have to be dropped here.

With this additional prescription, namely to shift the terms diverging
as the two velocities become equal into the states, we may as well obtain
a $1/m$ expansion of the short distance contribution by calculating
directly with (\ref{propstat}), i.e.~within the HQET framework. This
should coincide with what is obtained in using (\ref{propeff}) and
subsequent expansion in powers of $1/m$. The diverging terms reappear in
the states, since by an appropriate choice of the
leading order Lagrangian
(as in (\ref{lnull})) the corresponding terms are reproduced by the
residual momenta, but now they are buried in the non-perturbative
infrared physics of the states. The price to pay is a mass dependence
of the states, which is not accessible via a $1/m$ expansion.

This simplified prescription allows us to apply the
full machinery of HQET
to the calculation of the short-distance corrections in \HQQET,
and at the
one-loop level the results obtained for inclusive heavy quarkonia decays
coincide with what is obtained in NRQCD \cite{BB94}
and subsequent expansion
in powers of $1/m$ \cite{MS94}. A proof whether this
is true also in higher
orders in the loop expansion lies beyond the scope
of the present article.

The other type of corrections are the power corrections,
i.e.~the higher order
corrections in the $1/m$ expansion. These are induced by
the expansion of the
field and of the Lagrangian and are included as perturbations.
It has been noticed that the terms of order $1/m^2$ and higher of the
Lagrangian and the fields depend on the convention: As in any
effective theory it is possible to perform local field redefinitions,
thereby shifting certain terms from the Lagrangian into
the fields and vice
versa \cite{CC69}. However, any physical matrix element
remains unchanged,
if such a field redefinition is performed. For our purpose a convenient
definition of the higher order terms in the Lagrangian
is the one, in which
all terms, which would vanish by the use of the static equation of motion
derived from (\ref{lstat}), are shifted into the definition of the fields.
In this convention the Lagrangian up to and including $1/m^2$ terms takes
the form
\begin{equation}
{\cal L} = {\cal L}_{\mbox{static}} + {\cal L}_I
\end{equation}
where
\begin{eqnarray}
{\cal L}_{\mbox{static}} &=&
    \bar{h}^{(+)}\, ivD\, h^{(+)} -  \bar{h}^{(-)}\, ivD\, h^{(-)}
\nonumber\\
{\cal L}_I &=& \left(\frac{1}{2m} \right) L_1 +
               \left(\frac{1}{2m} \right)^2 L_2
\nonumber \\
           &=& \left(\frac{1}{2m} \right)
                     (K_1  +  G_1 ) +
                     \left(\frac{1}{2m} \right)^2
                     (K_2  + G_2 )  + {\cal O} (1/m^3)
\ .
\label{lexpFW}
\end{eqnarray}
Since now only a single velocity $v$ appears, we omit the
subscript $v$ from the field operators in the following.
Furthermore,  we have defined
\begin{eqnarray}
K_1 = K_1^{(+)} + K_1^{(-)} \quad
K_1^{(\pm)} &=&  \bar{h}^{(\pm)} [ (i D)^2 - (ivD)^2 ] h^{(\pm)}
\nonumber \\
G_1 = G_1^{(+)} + G_1^{(-)} \quad
G_1^{(\pm)} &=&  (-i) \bar{h}^{(\pm)} \sigma_{\mu \nu}
               (i D ^\mu)(i D^\nu)  h^{(\pm)}
\nonumber\\
K_2 = K_2^{(+)} + K_2^{(-)} \quad
K_2^{(\pm)} &=& \bar{h}^{(\pm)} [(i D_\mu) ,
[ (-ivD), (iD^\mu) ]] h^{(\pm)}
\nonumber \\
G_2 = G_2^{(+)} + G_2^{(-)} \quad
G_2^{(\pm)} &=&   (-i)  \bar{h}^{(\pm)}
\sigma_{\mu \nu} \{ (i D ^\mu) ,
                               [ (-ivD), (iD^\nu) ] \} h^{(\pm)}
\ .
\end{eqnarray}
The corresponding expansion of the field $\bar{Q}^{(+)}_v$ reads
\begin{eqnarray}
Q_v^{(+)}(x) &=& \left( 1 + \frac{1}{2m} P_-  i\fmslash{D}
              - \frac{1}{8m^2} (ivD) P_-  i\fmslash{D} \right.
\nonumber\\
&& \quad \left. - \frac{1}{8m^2} \left( (i D)^2 - (ivD)^2
                       - i \sigma_{\mu \nu} iD^\mu iD^\nu \right)
                       + {\cal O} (1/m^3)
                       \right) h^{(+)} (x)
\ .
\label{fexpFW}
\end{eqnarray}
In fact this is the form that has been obtained from QCD by a sequence
of Foldy-Wouthuysen transformations in \cite{KT91}.

The corrections to currents involving heavy quarks are then obtained
by inserting the expansion (\ref{fexpFW}) for the full fields of QCD,
while the corrections to the states are implemented via time-ordered
products of the expansion of the currents with higher-order terms of
the Lagrangian. However, here we have to take into account the fact
that the leading order dynamics of the states already contains the first
order kinetic energy contribution $K_1$ and hence this piece must not
to be included as a perturbation.

The higher order terms of the heavy mass  expansion for heavy quarkonia
are parametrized then by matrix elements
involving the ``unperturbed'' quarkonia states
that are described by ${\cal
 L}_0$
and operators expressed in terms of the static
fields $h_v^{(\pm)}$. Due to
the mass dependence of the states flavour symmetry is broken, which means
that these matrix elements are different for charmonium and
bottomonium states. These matrix elements are genuinely non-perturbative
quantities and spin symmetry may be used to count the number of
independent parameters. Accessing the actual values of these parameters
requires input beyond \HQQET; they may be extracted
from experiment, calculated
using some model framework, or eventually obtained from a lattice
calculation.

\section{Conclusions}
The heavy mass limit has been used with great success
for systems involving a
single heavy quark. In this limit the mass gap between
the particles and the
antiparticles becomes infinitely large and, as a consequence,
particle and
antiparticle numbers are separately conserved. HQET has been
formulated in the
one (anti)particle sector, and this is sufficient for almost
all applications
considered  so far.

In order to describe heavy quarkonia one has to deal
with the particle-antiparticle
sector. Treating both heavy constituents in the
static limit is only possible if
the two velocities $v$ and $v'$ are very different,
i.e~if $vv'-1 = {\cal O}(1)$. The naively
calculated QCD radiative corrections exhibit
logarithmically divergent and purely
imaginary contributions, which behave as $1/\sqrt{(vv')^2 - 1}$.
Interpreting these
pieces in the standard way as contributions to the
anomalous dimension yields then
phases, which are ill behaved as $v \to v'$.

It has been shown that these phases may be removed by a
redefinition of the
states \cite{KM93}. The phases are thus a property of the
states and and hence
an infrared effect. After redefinition of the states,
the ``true'' short distance
contributions, i.e~the anomalous dimensions become real.

In a heavy quarkonium state the velocities of the two heavy constituents
differ only by a small amount of the order $1/m$,
and it is appropriate to
choose the same velocity for both heavy particles,
which is then identified
with the velocity of the heavy quarkonium. However,
such a limit may not be taken
from the expression obtained in the static limit with $v \neq v'$.

In the present paper we have shown that one may
formulate a Heavy Quarkonium Effective
Theory (\HQQET), despite of these difficulties.
The key-point is that the leading order term
has to be chosen in such a way that the diverging
phases are generated by its dynamics.
This forced us to include also the first order
kinetic energy term into the leading
order Lagrangian (\ref{lnull}), which determines
the evolution of the states. Consequently
the ``unperturbed'' states have to depend on the
mass and heavy flavour symmetry is lost.
In other words, from this point of view it is
likely that there is no mass independent,
static limit of a heavy quarkonium state, since the
purely static Lagrangian does not
generate the diverging phases and probably also
does not generate bound states.

However, the Lagrangian (\ref{lnull}) still has an
unbroken heavy quark spin symmetry,
and hence the ``unperturbed'' heavy quarkonia have
to fall into spin symmetry quartets,
since all four orientations of the two heavy quark
spins will lead to degenerate states.
This symmetry also allows to reduce the number
of independent non-perturbative parameters.

This effective theory framework allows us also the
calculation of corrections.
The main result of the present analysis is that the
short distance corrections
may be calculated using the methods of HQET. However,
if the velocities of the
two heavy quarks become equal, divergent terms will
appear, which are infared
contributions and which have to be shifted into
the states. The appropriately
chosen ${\cal L}_0$  (\ref{lnull}) contains
these divergent terms as piece of its
infrared dynamics. The remaining expressions are
the ``true'' short distance
contributions, which remain well behaved as
$v \to v'$. This prescription has
been applied at the one-loop level and
yields the same $1/m$ expansion as a
NRQCD calculation with a subsequent
$1/m$ expansion.

The second type of corrections corresponds to the
$1/m$ expansion of the Lagrangian
and the fields. Including these contributions
proceeds along the same lines as in
HQET, with the only difference that no time-ordered
products with the first order
kinetic energy operator $K_1$ are present,
since this is included already in the
definition of the ``unperturbed'' states.

\HQQET{} may be applied to all processes in
which the two heavy particles
annihilate, such as inclusive and exclusive
decays of charmonium and bottomonium,
(electromagnetic and strong) and also to
weak decays of $B_c$. Some applications
of this idea have been investigated in
\cite{MS94} and using related methods in
\cite{BB94}, but \HQQET{} opens a wide field
of further applications.

\section*{Acknowledgements}
We want to thank T.\ Ohl for clarifying discussions.

\end{document}